\documentclass[hyper,a4paper]{JHEP3} 
\usepackage{graphicx}

\newcommand{\be}{\begin{equation}}
\newcommand{\ee}{\end{equation}}
\newcommand{\bea}{\begin{eqnarray}}
\newcommand{\eea}{\end{eqnarray}}
\newcommand{\no}{\nonumber}
\newcommand{\cM}{{\cal M}}

\newcommand{\cO}{{\cal O}}
\newcommand{\gvsg}{g^{_V}_{_{S \gamma}}}
\newcommand{\gphif}{g^{_\Phi}_{_{f_0 \gamma}}}
\newcommand{\gvpg}{g^{_{V}}_{_{P \gamma}}}
\newcommand{\gvpgd}{g^{_{V^\prime}}_{_{P \gamma}}}
\newcommand{\gvpv}{g^{_V}_{_{P V^\prime}}}
\newcommand{\gs}{g^{_S}_{_{12}}}

\newcommand{\gpp}{g^{_{f_0}}_{_{\pi\pi}}}
\newcommand{\veps}{\varepsilon}
\newcommand{\im}{\mbox{Im}}
\newcommand{\re}{\mbox{Re}}

\title{The $e^+ e^- \to P_1 P_2 \gamma$ processes close to the $\Phi$ peak:
toward a model-independent analysis}

\author{Gino Isidori~${}^{a}$, Luciano Maiani~${}^{b}$,  
 Maria Nicolaci~${}^{b}$, Simone Pacetti~${}^{a}$ \\
${}^{a}~$INFN, Laboratori Nazionali di Frascati, I-00044 Frascati,  Italy  \\ 
${}^{b}~$Dipartimento di Fisica, Universit\`a di Roma ``La Sapienza''  \\
${}^{~}~$INFN, Sezione di Roma, P.le A. Moro 2, I-00185 Roma, Italy \\
E-mail: 
\email{gino.isidori@lnf.infn.it}, 
\email{luciano.maiani@roma1.infn.it},
\email{maria.nicolaci@roma1.infn.it},
\email{simone.pacetti@lnf.infn.it} }

\preprint{\hepph{0603241}}	

\abstract{We discuss the general decomposition and possible general 
parameterizations of the processes $e^+ e^- \to \gamma^* 
\to P_1 P_2 \gamma$, where  $P_1 P_2=\pi^0 \pi^0$,
$\pi^0\eta$, or $\pi^+\pi^-$, for $\sqrt{s}\approx M_\Phi$.   
Particular attention is devoted to the amplitude where 
the two pseudoscalar mesons are in a $J^{CP}= 0^{++}$ state,
where we propose a general parameterization 
which should help to shed light on the 
nature of light scalar mesons.}


\begin{document}
          
\section{Introduction}

As widely discussed in the literature,
radiative $\Phi$ decays, such as  
$\Phi \to f_0 \gamma \to \pi\pi \gamma$
or $\Phi \to a_0 \gamma \to \pi\eta \gamma$,
are one of the primary sources of information about 
the interesting and still controversial sector 
of light scalar mesons (see e.g.~Refs.~\cite{handbook,vari,MPPR}
and references therein). In principle, the large 
amount of data collected at $\Phi$ factories should 
allow to study these processes with excellent accuracy. 
However, it must be realized that these 
processes are only one of the components of the
basic $\Phi$-factory observables, namely the 
$e^+ e^- \to P_1 P_2 \gamma$ cross sections
(where  $P_1 P_2=\pi^0 \pi^0$, $\pi^0\eta$, or $\pi^+\pi^-$).
An accurate and possibly model-independent 
description of all the components of these reactions
is necessary in order to extract reliable 
information about the scalar sector of QCD.

The purpose of the present paper is twofold.
First, we discuss how to isolate the 
interesting scalar amplitude from the other 
contributions to the cross section. Second, 
we present a general parameterization of the 
scalar amplitude which should allow a model-independent 
determination of basic parameters 
such as masses, widths and couplings of $f_0$ 
and $a_0$ mesons.

The paper is organized as follows. In Section~\ref{sect:general}
we discuss the general decomposition the cross sections
in terms of gauge-invariant tensors. We start from the
simpler case of  neutral final states, $|\pi^0\pi^0\gamma\rangle$ and $|\pi^0\eta\gamma\rangle$.
We then generalize to the  $|\pi^+\pi^- \gamma\rangle$ case,
where initial- and final-state radiation represent a 
serious background. In Section~\ref{sect:vector} we analyse the 
contributions to the invariant amplitudes due to the exchange 
of vector mesons ($e^+e^- \to V P_{1(2)} \to P_1 P_2 \gamma $), 
which represent an irreducible physical background for the scalar amplitude. 
Finally, in Section~\ref{sect:scalar}
we present a general decomposition of the 
scalar amplitude ($e^+e^- \to S \gamma \to P_1 P_2 \gamma$)
which takes into account the 
narrow-width structure of $f_0$ and $a_0$ at high $M_{ P_1 P_2}$,
and the  constrains of unitarity 
and chiral symmetry at low $M_{ P_1 P_2}$.
The complete procedure we propose for the  
fit of the cross sections 
is summarized in the last section. 

\section{General decomposition of the cross sections}
\label{sect:general}

\subsection{Neutral final states}
The generic matrix element for the transition $e^+e^- \to P_1^0 P_2^0 \gamma$ 
($P_{1,2} = \pi^0$ or $\eta$) can be written as 
\be
\cM\left[ e^+(p_+) e^-(p_-) \to P_1^0(p_1)  P_2^0(p_2)  \gamma( \veps,k) \right] 
= \frac{e}{s} \bar v(p_+)\gamma_\mu u(p_-) T^{\mu\nu}  \veps_\nu~,
\label{eq:Mdec}
\ee
where $s = (p_+ + p_-)^2 \equiv P^2$. The constraints of gauge 
invariance imply
\be
P_\mu T^{\mu \nu}=k_\nu T^{\mu \nu} = 0~.
\ee
Using this notation, 
the differential cross section with 
unpolarized beams becomes
\bea
d\sigma(e^+e^- \to P_1^0  P_2^0 \gamma) &=& \frac{ 1 }{ 8  s} C_{12}\sum_{\rm spins}   
\left| \cM \right|^2  d\Phi \no \\
 &=&  \frac{ 8 \pi \alpha  }{  s^3} C_{12} \left[ p_+^\mu p_-^\nu + p_-^\mu p_+^\nu - 
\frac{s}{2} g^{\mu\nu} \right]
\left[ -\frac{1}{4} g^{\rho \sigma} T_{\mu \sigma} T^*_{\nu \rho} \right] d\Phi~,
\eea
where 
\be
d\Phi = \frac{d^3 k}{(2\pi)^32 E_\gamma} 
\frac{d^3 p_1 }{(2\pi)^32E_1}\frac{d^3 p_2}{(2\pi)^32 E_2}  (2\pi)^4 
\delta^{(4)}(P-p_1-p_2-k)
\ee
and $C_{12}$ takes into account the 1/2 factor in case of identical 
particles: $C_{12}=1$ for $\{P^0_1,P_2^0\}\equiv \{\eta,\pi^0 \}$; 
$C_{12}=1/2$ for $\{P^0_1,P_2^0\}=\{\pi^0,\pi^0\}$.

As a result of the gauge-invariance conditions, in the case of 
a real photon in the final state we can decompose $T^{\mu \nu}$ as the sum 
of three independent structures:
\be
T_{\mu \nu}= \frac{4}{s} \left[  A_1 L^{(1)}_{\mu \nu} + A_2 L^{(2)}_{\mu \nu} 
                          + A_3 L^{(3)}_{\mu \nu}  \right]~,
\label{eq:Tdec}
\ee
with
\bea 
&& L^{(1)}_{\mu \nu}=(k\cdot P)g_{\mu \nu}-k_\mu P_\nu~,   \no \\ 
&& L^{(2)}_{\mu \nu}= \frac{4}{s} \left\{  (P \cdot q) \left[(k\cdot q) g_{\mu\nu} - k_\mu q_\nu \right]
+(k\cdot P) q_\mu q_\nu -(k\cdot q) q_\mu P_\nu  \right\}~, \no \\ 
&& L^{(3)}_{\mu \nu}=\frac{4}{s} \left[ (k\cdot q)(P^2g_{\mu \nu}-P_\mu P_\nu) +
 (k\cdot P)P_\mu q_\nu -P^2 k_\mu q_\nu \right]~,  
\qquad\quad  q=\frac{1}{2}(p_1-p_2)~. \no 
\eea
The normalization of $T_{\mu \nu}$ and the $L^{(i)}_{\mu \nu}$ has been
chosen in order to maximize the contact with Ref.~\cite{vepp}: 
$L^{(1)}_{\mu \nu}$ and $L^{(3)}_{\mu \nu}$ are identical
to the corresponding expressions of Ref.~\cite{vepp}, 
while $L^{(2)}_{\mu \nu}$ coincides with the corresponding tensor 
of Ref.~\cite{vepp} only in the case of identical particles. 

The dimensionless form factors $A_i$ are determined by the 
specific dynamical model. For instance, the scalar amplitude 
$e^+e^- \to S \gamma \to P_1 P_2\gamma$ induces a non-vanishing 
contribution only to $A_1$, while the transition  
$e^+e^- \to V P_{1(2)} \to P_1 P_2 \gamma $ leads to 
non-vanishing contributions to all the form factors. 

The phase space element can be re-written as
\be
d\Phi = \frac{1}{8 (2\pi)^4} d E_1 dE_\gamma d\Omega_{\rm b} = \frac{1}{8 (2\pi)^4} d E_1 dE_2 d\Omega_{\rm b} ~,
\ee
where $E_{\gamma,1,2}$ are the energies 
in the c.o.m. frame and $\Omega_{\rm b}$ denotes 
the solid angle of the beam axis with respect to the decay 
plane.\footnote{~The element of the solid angle can be expressed 
as $d\Omega_{\rm b}=d\cos\theta_{\rm \gamma} d\phi$, where $\theta_{\rm \gamma}$
is the angle between photon and $e^+$ momenta, and $\phi$ 
the orthogonal angle with respect to the decay plane. 
The latter leads to a trivial integration in the $|A_1|^2$ term, 
but is non-trivial for the other contributions. }
Performing the angular integration we find
\bea
F(x,x_1,x_2) &=& \frac{1}{2\pi s} \int d \Omega_{\rm b}  \left[
-\frac{1}{4} g^{\rho \sigma} T_{\mu \sigma} T^*_{\nu \rho} \right] 
\left[ p_+^\mu p_-^\nu + p_-^\mu p_+^\nu - \frac{s}{2} g^{\mu\nu} \right]
\no \\ 
&=& a_{11}|A_1|^2+a_{22}|A_2|^2+a_{33}|A_3|^2+ 
a_{12}(A_1A_2^*+A_1^*A_2) \no \\ && + a_{13}(A_1A_3^*+A_1^*A_3)+
a_{23}(A_2A_3^*+A_2^*A_3)~,  \label{evme} \end{eqnarray} \noindent
where the $a_{ij}$ coefficients are given by 
\bea 
a_{11} &=& \frac{4}{3}x^2 \no \\ 
a_{12} &=& \frac{2}{3} \left[  (x_1-x_2)^2 + x^2(\sigma-1+x) 
                              - 2 \delta(x_1-x_2) +\delta^2 \right]  \no \\ 
a_{13} &=& \frac{8}{3} x(x_1-x_2 -\delta) \no \\ 
a_{22} &=& \frac{1}{3}\left\{ (x_1-x_2)^4+ 2 x^2 ( \sigma- 1+x )^2 
           +2(x_1-x_2)^2(1-x)( \sigma-1+x ) \right.  \no \\
&&  - \delta(x_1-x_2)\left[ 2(x_1-x_2)^2 + 2 (\sigma-1+x)(x_1+x_2) \right] \no \\
&& \left. +\delta^2\left[ (x_1-x_2)^2 + 2(\sigma-1+x) \right] \right\} \no \\ 
a_{23} &=& \frac{4}{3}(x_1-x_2)\left[ (x_1-x_2)^2 - 2 \delta(x_1-x_2) +\delta^2 \right] \no \\ 
a_{33} &=& \frac{8}{3}\left[ (x_1-x_2)^2(1+x)-x^2( \sigma-
1+x )  -\delta(x_1-x_2)(x+2)+\delta^2 \right]  \label{cij} 
\eea
in terms of the adimensional variables 
\bea 
&& x_i = \frac{2 E_i}{\sqrt{s} } \qquad\qquad  
x = \frac{2 E_\gamma }{\sqrt{s}} = 2-x_1-x_2  
 \no \\
&&      \sigma = \frac{2 (M^2_{1} +  M^2_{2}) }{s} \qquad\qquad   
        \delta = \frac{2 (M^2_{1} -  M^2_{2}) }{s}.
\eea  
In the limit $\delta \to 0$ the $a_{ij}$ coefficients in Eq.~(\ref{cij}) 
coincide with the $C_{ij}$ of Ref.~\cite{vepp}, with the exception of 
$a_{12}$, which is reported incorrectly in Ref.~\cite{vepp}. 

The general expression for the total cross-section 
is then given by 
\bea
d \sigma(e^+e^-\to P_1^0 P_2^0 \gamma) &=& \frac{\alpha }{ 32 \pi^2 s  } C_{12}
~F(2-x_1-x_2,x_1,x_2)~d x_1 dx_2\no \\ 
&=& \frac{\alpha}{ 32 \pi^2 s  } C_{12} 
~F(x,x_1,2-x-x_1) ~d x~ dx_1 .
\label{eq:general}
\eea

\subsection{Initial- and final-state radiation and generalization 
to the $e^+e^- \to \pi^+ \pi^- \gamma$ case}
  
At $\cO(\alpha^2)$  we can distinguish three basic 
components to the processes $e^+e^- \to \pi^+ \pi^- \gamma$: 
initial-state radiation (ISR), final-state radiation (FSR)
and the irreducible structure-dependent (SD) amplitude
which vanishes in the  $E_\gamma \to 0$ limit 
(see Fig.~\ref{fig:amplitudes}). The latter, which is identical 
to the neutral case discussed before, is the only contribution 
sensitive to scalar-meson dynamics.

\FIGURE[t]{
\includegraphics[scale=0.80]{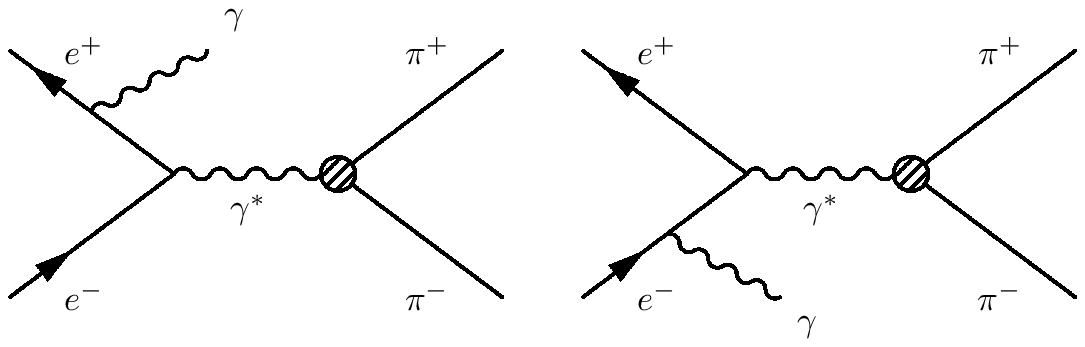}\\[-5pt]
\centerline{\footnotesize  a) [ISR] \hskip 3.5 cm  b) [ISR] \hskip 1.5 cm }\\[15pt] 
\includegraphics[scale=0.80]{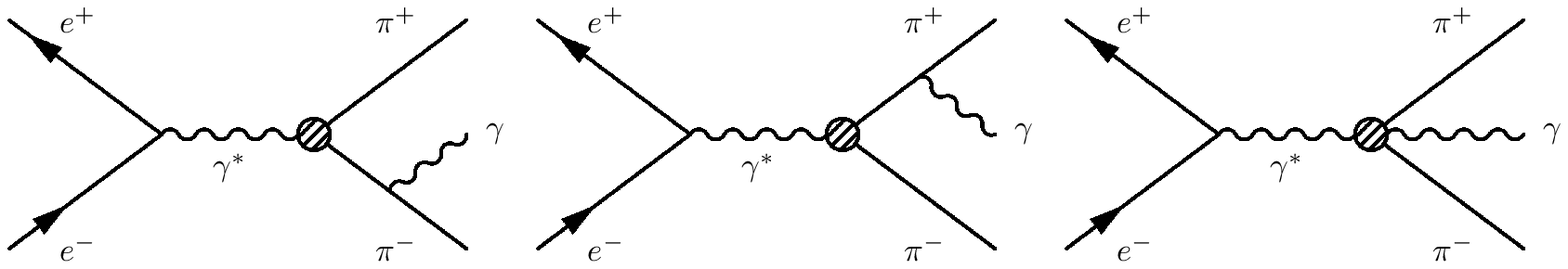}\\[-5pt]
\centerline{\footnotesize  c) [FSR] \hskip 3 cm  d) [FSR] \hskip 3 cm  e) [FSR+SD] }
\caption{\label{fig:amplitudes}
Tree-level diagrams for ISR, FSR  and SD
contributions to $e^+e^- \to \pi^+ \pi^- \gamma$.
The SD amplitude corresponds to the $\cO(E_\gamma)$ 
gauge-invariant terms in the diagram e).}
}

By construction, ISR and FSR amplitudes can be fully 
described in terms of the electromagnetic 
form factor of the pion and are known with good 
accuracy\footnote{~We denote here 
as FSR amplitude only the gauge-invariant part 
of the amplitude which can unambiguously related to the on-shell 
non-radiative $e^+e^- \to \pi^+ \pi^-$ processes, according 
to Low's theorem~\cite{Low}. }  (for an extensive discussion applied to the 
$\Phi$-factory case see e.g.~Ref.~\cite{ISR,Pancheri}).
The $|\pi^+ \pi^- \rangle$ state produced 
by ISR has opposite transformation properties 
under parity and  charge conjugation
with respect to those produced by FSR and SD.
For this reason, as long as the kinematical cuts 
applied on the cross section are symmetric under 
the exchange $\pi^+ \leftrightarrow \pi^-$, we can 
neglect the interference of the ISR amplitude 
with the other two contributions.
On the other hand, the FSR amplitude, which can be 
decomposed according to Eq.~(\ref{eq:Tdec}), 
does interfere with the SD terms:
\be
d \sigma (e^+e^- \to \pi^+ \pi^- \gamma)
 = d \sigma_{\rm ISR}  + d \sigma_{\rm FSR} +
d \sigma_{\rm SD}  + d \sigma^{\rm Interf}_{\rm FSR+SD}~.
\label{eq:sigma_tot}
\ee
The form factors of the FSR contribution, which are singular
in the $x\to 0$ limit, can be written as
\be
A_1^{\rm FSR} = -  \frac{ 4\pi\alpha x  F_{\pi\pi}(s) }{x^2-(x_1-x_2)^2}~,
\qquad 
A_2^{\rm FSR} =  \frac{8 \pi\alpha F_{\pi\pi}(s)}{x^2-(x_1-x_2)^2}~,
\qquad 
A_3^{\rm FSR} = 0~,
\label{eq:Ai_FSR}
\ee
where $F_{\pi\pi}(s)$ denotes the pion electromagnetic 
form factor (with the standard normalization $F_{\pi\pi}(0)=1$).
The last three terms in Eq.~(\ref{eq:sigma_tot}) can thus 
be described by a generic expression of the type in Eq.~(\ref{eq:general}),
where the form factors are obtained by summing
the singular $A_i^{\rm FSR}$ to the regular terms 
of the SD amplitude. 

\medskip 

So far we have analysed only the $\cO(\alpha^2)$
contributions to the cross sections, or the leading 
contributions in the case of a single detected photon in the final sate. 
As is well know, the physical cross sections are obtained by the 
convolution of these leading expressions with appropriate radiation 
functions \cite{ISR_russi}
which take into account the effect of the undetected 
ISR (both for charged and neutral final states) and the undetected 
FSR (for the charged final state only).

\FIGURE[t]{
\includegraphics[scale=1.0]{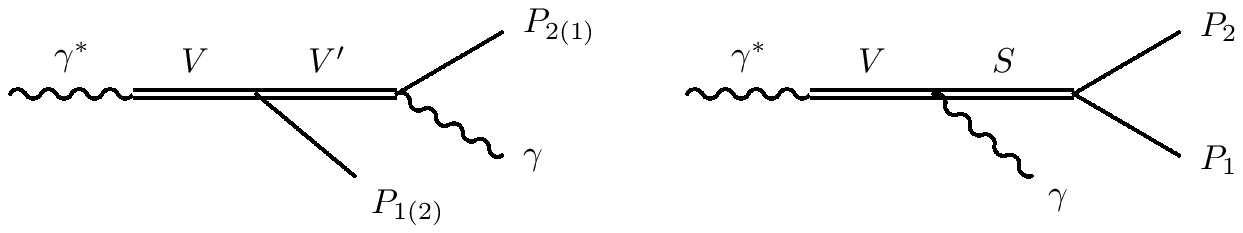}
\caption{\label{fig:vv} Leading contributions to the 
structure-dependent amplitude: vector-meson exchange 
(left) and scalar (right) contributions. }}

\section{The vector-meson exchange contribution}
\label{sect:vector}
In a good fraction of the allowed kinematical region, 
a sizable contribution to the structure-dependent amplitude
is induced by the vector-meson exchange process in Fig.~\ref{fig:vv} (left).

The starting point to evaluate this contribution 
are the effective couplings $\gvpv$ and $\gvpg$
(with dimension 1/energy), defined by  
\bea
\cM\left[ V(\tilde \veps, P) \to P(q) \gamma(\veps,k) \right] &=& e \gvpg~
\epsilon^{\mu\nu\rho\sigma} \tilde \veps_\mu \veps_\nu q_\rho k_\sigma~,  \no \\
\cM\left[ V(\tilde \veps, P) \to P(q) V^\prime (\veps,k) \right] &=& \gvpv~
\epsilon^{\mu\nu\rho\sigma} \tilde \veps_\mu \veps_\nu q_\rho k_\sigma~,
\label{eq:gvmd}
\eea
and the adimensional vector-meson electromagnetic 
couplings, $F_V$, defined by 
\be
\cM\left[ V(\veps) \to e^+ e^- \right] = \frac{e}{F_V} ~ \veps_\mu ~
 \bar u(p_-)\gamma^\mu v(p_+)~.
\label{eq:FV}
\ee
In terms of these couplings, the vector-meson exchange process in  Fig.~\ref{fig:vv}
give rise to the following contributions to the form factors:
\bea
A^{\rm vect}_1 &=& -\frac{1}{4} \left( -1+\frac{3}{2}x+\sigma\right)\left[g(x_1)+g(x_2)\right] 
 +\frac{1}{4}\left( x_1-x_2 +\frac{1}{2} \delta \right)[g(x_1)-g(x_2)] \no \\
A^{\rm vect}_2 &=& \frac{1}{4}\left[g(x_1)+g(x_2)\right] \no \\
A^{\rm vect}_3 &=& -\frac{1}{8}\left[g(x_1)-g(x_2)\right] 
\label{eq:A_vect}
\eea
where 
\be
g(y) = \sum \limits_{V,V^\prime=\rho,\omega,\Phi,\rho^\prime,\omega^\prime,\ldots} 
\frac{ e \gvpv \gvpgd }{4 F_V} ~\frac{ s^2 M_{V}^2  }{ D_V(s) D_{V^\prime} [(1-y)s]}
\ee
and 
\be
D_X(q^2) = s - M_X^2  +i M_X \Gamma_X~.
\label{eq:DX}
\ee
In the limit of identical particles in the final state, 
these results are fully consistent with those of Ref.~\cite{vepp}.
As discussed in Ref.~\cite{vepp}, 
a natural improvement of the above expressions 
is obtained by the replacement of the constant 
$\Gamma_X$ with appropriate energy-dependent widths, 
taking into account the velocity 
factors of the dominant final states \cite{flatte}.

\TABLE[t]{
\begin{tabular}{|l|l|l|}
\hline 
 \multicolumn{1}{|c|}{\raisebox{3pt}[16pt][0pt]{$F_V$}} & 
 \multicolumn{1}{c|}{\raisebox{3pt}[16pt][0pt]{$\gvpg~\left[{\rm GeV}^{-1}\right]$}} & 
 \multicolumn{1}{c|}{\raisebox{3pt}[16pt][0pt]{$\gvpv = 
   g^{_{V^\prime}}_{_{P V}}~\left[{\rm GeV}^{-1}\right]$}} 
\\ \hline\hline
 \raisebox{2pt}[14pt][0pt]{ }  & 
 \raisebox{2pt}[14pt][0pt]{$|g^{_{\rho}}_{_{\pi \gamma}}  |= 0.79 \pm 0.10$}  &
 \raisebox{2pt}[14pt][0pt]{$|g^{_{\omega}}_{_{\pi \rho}}|=  12   \pm 2 $   }
\\ 
 \raisebox{10pt}[14pt][0pt]{$|F_\rho |= 16.6 \pm 0.2$}  & 
 \raisebox{2pt}[14pt][0pt]{$|g^{_{\rho}}_{_{\eta \gamma}} |= 1.56 \pm 0.08$}  &
 \raisebox{2pt}[14pt][0pt]{$|g^{_{\Phi}}_{_{\pi \rho}} |= 1.3 \pm 0.2$}
\\ \cline{1-2}
 \raisebox{2pt}[14pt][0pt]{ } & 
 \raisebox{2pt}[14pt][0pt]{$|g^{_{\omega}}_{_{\pi \gamma}}  |= 2.38 \pm 0.04$} &  
 \raisebox{2pt}[14pt][0pt]{$|g^{_{\Phi}}_{_{\pi \omega}} |= (4.1 \pm 0.5) \times 10^{-2}$ } 
\\  
 \raisebox{10pt}[14pt][0pt]{$|F_\omega |= 55.9 \pm 0.5$} & 
 \raisebox{2pt}[14pt][0pt]{$|g^{_{\rho}}_{_{\eta \gamma}}   |= 0.46 \pm 0.02$} &
 \raisebox{2pt}[14pt][0pt]{$|g^{_{\Phi}}_{_{\eta \omega}} |= 1.1 \pm 0.2$}
\\  \cline{1-2}
 \raisebox{2pt}[14pt][0pt]{ } & 
 \raisebox{2pt}[14pt][0pt]{$|g^{_{\Phi}}_{_{\pi \gamma}}  |= 0.132 \pm 0.004$ }  &
 \raisebox{2pt}[14pt][0pt]{$|g^{_{\Phi}}_{_{\eta \Phi}}|= 2|g^{_{\rho}}_{_{\eta \rho}} |= 
   2|g^{_{\omega}}_{_{\eta \omega}} |=  7.0  \pm 1.5$ } 
\\  
 \raisebox{10pt}[14pt][0pt]{$|F_\Phi |= 44.4 \pm 0.4 $} & 
 \raisebox{2pt}[14pt][0pt]{$|g^{_{\Phi}}_{_{\eta \gamma}}  |= 0.691 \pm 0.008$ }  &
 \raisebox{2pt}[14pt][0pt]{ } 
\\  \hline
\end{tabular}
\caption{\label{tab:vmd} 
Reference values for the modulos of the effective vector-meson couplings 
defined in Eqs.~(\ref{eq:gvmd})--(\ref{eq:FV}).} }

In Table~\ref{tab:vmd} we report the current estimates for the 
most relevant set of  $F_V$,  $\gvpg$, and $\gvpv$ couplings.
The results for $F_V$ 
and  $\gvpg$ have been determined by means of the 
relations 
\be
 \Gamma(V \to e^+ e^-) = \frac{ \alpha M_V }{3 \left| F_V\right|^2 }~, \qquad
 \Gamma(V \to P \gamma ) = \frac{ \alpha |\gvpg|^2 }{3}   
 \left[\frac{ M_V^2 -M_P^2}{2 M_V}\right]^3~,  
\ee
using the experimental values 
of $\Gamma(V \to e^+ e^-)$ and $\Gamma(V \to P \gamma )$
in \cite{PDG}. The $\gvpv$ have been determined 
theoretically,\footnote{~The $\gvpv$ have been determined by: 
i)~assuming a simple effective 
Lagrangian of the type ${\cal L}~= ~g~{\rm tr}(\{ V, V^\prime \} P)$, 
where $V$, $ V^\prime$, and $P$ are $3\times 3$ matrices in flavour space;
ii)~fixing the $\Phi$--$\omega$ mixing angle from the $F_V$ values; 
iii)~enforcing the vector-meson dominance relation 
$\gvpg = \sum_{V^\prime} \gvpv/(e  F_{V^\prime} )$.}
with the exception of $g^{_{\Phi}}_{_{\pi \omega}}$, 
which has been determined directly from the experimental 
value of $\Gamma(\Phi \to \omega \pi)$ in \cite{PDG}. 
We stress that the results reported in Table~\ref{tab:vmd} should 
be considered only as rough reference values, or as  natural starting 
point for a fit of the cross sections.
The high-statistics data on the $e^+ e^- \to P_1 P_2 \gamma$ 
reactions at a $\Phi$ factory should allow to determine 
these couplings (or at least some combinations of them)
with much higher accuracy.

\section{The scalar amplitude}
\label{sect:scalar}
We are now ready to analyse the scalar amplitude or, 
more precisely, the contributions to the form factor $A_1$
not described by vector mesons and/or FSR.  

The contribution to $A_1$ induced by the exchange of a 
single vector and a single scalar resonance, as shown in Fig.~\ref{fig:vv} (right), is:
\be
A_1(e^+e^- \to V \to S \gamma \to P_1 P_2\gamma)
 = \frac{ e \gs \gvsg}{4 F_V} \frac{ s M_V^2}{D_V(s) D_S[(1-x)s]}   
\ee
where the couplings $\gs$ and  $\gvsg$ are defined by
\bea
\Gamma(V \to S \gamma) &=& \frac{ \alpha |\gvsg|^2}{3} 
 \left[\frac{ M_V^2 -M_S^2}{2 M_V}\right]^3~,  \\
\Gamma(S \to P_1 P_2) &=&  \frac{ |\gs|^2~ p_{12}^*(M^2_S)}{8 \pi M^2_S}~,\label{eq:gamma12}  \\
p_{12}^*(M^2)  &=& 
\frac{ \left[M^2 -(M_1-M_2)^2\right]^{1/2}\left[M^2 -(M_1+M_2)^2\right]^{1/2} }{2 M}~. \no
\eea
If we could consider only this resonant contribution,
the total cross-section would assume the following form 
\bea
d \sigma_{\rm Scalar} &=& 
\frac{2 \alpha}{ 3 \pi^2 s^3} C_{12}  
\left| A_1(e^+e^- \to V \to S \gamma \to P_1 P_2\gamma) \right|^2  E_\gamma^2 
d E_\gamma d E_1  \no\\  &=&  \frac{C_{12} }{4\pi^2s} 
  BW_V(s) \frac{ \Gamma(V \to e^+e^-)}{M_V \Gamma^2_V}  \left|
  \frac{  e \gs \gvsg   }{ D_S(s_{12})  }   \right|^2 
 ~ \frac{  M_V^2 E^3_\gamma ~ p_{12}^*(s_{12}) }{ \sqrt{s_{12}} } d E_\gamma~,  \qquad\
\label{eq:sigma_SS}
\eea
where 
\bea
\quad s_{12} &=& s - 2\sqrt{s} E_\gamma = (1-x)s~,  \no \\
BW_V(s) &=& \frac{M^2_V \Gamma^2_V }{ |D_V(s)|^2 }= \frac{M^2_V \Gamma^2_V }{(M_V^2-s)^2 
+ M^2_V \Gamma^2_V }~.
\eea
Note that the $E_\gamma^3$ factor in (\ref{eq:sigma_SS}),
which is dictated by gauge-invariance according 
to the general decomposition (\ref{cij}),
implies a sizable distortion from a standard 
Breit-Wigner shape for resonances 
close to the end of the phase space,
such as $f_0(980)$ and $a_0(980)$.

The simplified expression (\ref{eq:sigma_SS}) is often used in the literature 
to describe the contributions 
of the narrow resonances $f_0(980)$ and $a_0(980)$. However, a coherent 
description of all the amplitudes contributing to the physical processes 
requires a more refined treatment. 
First, in order to compute the total cross section 
we need to consider the general expression in Eq.~(\ref{eq:general}),
with coherent sum of all contributions to the $A_i$:
\be
  A_1^{\rm full} = A_1^{\rm FSR} + A_1^{\rm vect} + A_1^{\rm scal}~, \qquad 
  A_2^{\rm full} = A_2^{\rm FSR} + A_2^{\rm vect}~, \qquad 
  A_3^{\rm full} = A_3^{\rm vect}~, 
\label{eq:A_full}
\ee
with   $A_i^{\rm FSR}$ ($\pi^+\pi^-$ case only) and $A_i^{\rm vect}$  
given in  Eq.~(\ref{eq:Ai_FSR}) and Eq.~(\ref{eq:A_vect}), respectively. 
Second, the expression of $A_1^{\rm scal}$ can involve 
several resonances and, possibly, also non-resonant backgrounds.

Since $f_0(980)$ and $a_0(980)$ are the only two narrow scalar 
resonances with mass below $M_\Phi$, a convenient parameterization 
for $A_1^{\rm scal}$ is 
\be
A_1^{\rm scal}  = \sum  \limits_{V=\rho,\omega,\phi;~ S=f_0,a_0} 
\frac{ e }{4 F_V} \frac{ s M_V^2}{D_V(s)}\left[ \frac{ \gs \gvsg }{ D_S(s_{12})} 
+  R^V_{12}(s_{12}) 
\right]~.
\label{eq:A1_gen}
\ee
Here $R^V_{12}(s_{12})$ denotes a non-resonant 
term which, in absence of re-scattering effects,
can be expanded as a regular series in powers of $(s_{12}-M_S^2)$. 
The situation becomes particularly simple in  the exact 
isospin limit, where a single narrow resonance 
can contribute to each channel: the $f_0(I=0)$ 
in the two $|\pi\pi\rangle$ cases and the $a_0(I=1)$
in the $|\pi\eta\rangle$ one. In this limit, at fixed values 
of $s$ and neglecting re-scattering effects, we expect 
a structure of the type
\be
A_1^{\rm scal}  \propto
  \frac{ g^S_{\rm eff} }{ s_{12} - M_{S}^2 + i M_{S} \Gamma_{S} }
+ \frac{\alpha_0}{M_\Phi^2} +\frac{\alpha_1}{M_\Phi^4} ( s_{12} -  M_{S}^2)
+ \cO[ ( s_{12} -  M_{S}^2)^{2} ]~,
\label{eq:par1}
\ee
with a different set of effective couplings for each channel.
As we will discuss in the following, this structure 
can be systematically improved in order to take into 
account the elastic re-scattering phases 
of the two pseudoscalar mesons.

It is worth to stress that a sizable non-resonant term 
is phenomenologically required from data, at least 
in the $|\pi\pi\rangle$ channels. Indeed, if we only retain 
the pole term in Eq.~(\ref{eq:par1}), the $E_\gamma^3$ factor 
in the cross section --see Eq.~(\ref{eq:sigma_SS})-- 
implies a too large result at low $s_{\pi\pi}$ compared to 
observations (see e.g.~Ref.\cite{KLOEp0p0,KLOEpp}):
experimental data clearly indicate that at low $s_{\pi\pi}$ 
the contribution of the $f_0(980)$ is partially 
compensated by other contributions.
 
In the $|\pi\pi\rangle$ channels one can consider 
an alternative parameterization where, in addition to the 
narrow $f_0(980)$, also the broad $f_0(600)$ 
is included by means of an appropriate complex propagator in the 
$I=0$ channel.\footnote{~The existence of a pole 
in the $S$-wave, $I=0$, $\pi\pi$ scattering amplitude 
--corresponding to the $f_0(600)$-- is not under doubt~\cite{Caprini}. 
However, this pole is very far from the real axis and 
quite close to the  $\pi\pi$ threshold~\cite{Caprini}.
As a consequence, its contribution to the amplitude (\ref{eq:A1_gen})
is not necessarily well described by a simple
complex propagator as in the $f_0(980)$ case.}
In this case the $f_0(600)$ could 
be responsible for the partial cancellation of 
the $f_0(980)$ contribution at low $s_{\pi\pi}$, with a minor  
role played by the non-resonant term.
In principle, precise enough experimental 
data on the $e^+ e^- \to \pi\pi \gamma$ cross 
sections should be able to distinguish 
the case of an explicit pole structure 
for the $f_0(600)$ from a pure polynomial term.
However, the broad nature of the  $f_0(600)$ makes 
this distinction very difficult, even with the high 
statistics available at a $\Phi$ factory.

\subsection{Re-scattering phases for $|\pi\pi\rangle$ final states}
\label{sect:rescattering}
As anticipated, the parameterization (\ref{eq:par1}) can be 
improved in order to take into account the absorptive parts
generated by elastic re-scattering. We illustrate here how this 
improvement can be implemented in the $|\pi\pi\rangle$  case, where this effect 
is more relevant and the information about elastic re-scattering 
at low $s_{\pi\pi}$ is very precise. For simplicity, we consider  
only $\Phi$-mediated contributions and we include only the 
$f_0(980)$ as explicit pole structure. 

Neglecting non-$\Phi$ contributions, we can  decompose the scalar 
form factors for charged and neutral $|\pi\pi\rangle$ final states as 
\bea
A_{1}^{\rm scal} (\pi^+\pi^-) &=& \frac{e}{4F_\Phi}\frac{sM_\Phi^2}{D_\Phi(s)}
~\frac{F^{\rm scal}_2(s_{\pi\pi})+\sqrt{2}F^{\rm scal}_0(s_{\pi\pi})}{\sqrt{3}}~, \no\\
A_{1}^{\rm scal} (\pi^0\pi^0) &=& \frac{e}{4F_\Phi}\frac{sM_\Phi^2}{D_\Phi(s)}
~\frac{\sqrt{2}F^{\rm scal}_2(s_{\pi\pi})-F^{\rm scal}_0(s_{\pi\pi})}{\sqrt{3}}~,
\label{iso-dec}
\eea
where $s_{\pi\pi}=(1-x)s$ and 
the reduced $F^{\rm scal}_{0,2}$ correspond the $I=0,2$ isospin combinations.
If we include only the $f_0(980)$ as explicit pole structure, we then 
have\footnote{~The effective coupling $\gpp$ denotes the coupling of the 
$f_0(980)$ to the two-pion state with $I=0$. Correspondingly, 
the decay width appearing in the definition (\ref{eq:gamma12})
must be interpreted as  $\Gamma[f_0 \to (\pi\pi)_{I=0}]=\Gamma(f_0 \to \pi^+\pi^-)
+\Gamma(f_0 \to \pi^0\pi^0)$.}
\bea
F^{\rm scal}_0(s_{\pi\pi}) &=&
\frac{ \gpp \gphif}{ s_{\pi\pi} - M_{f_0}^2 + i\sqrt{s_{\pi\pi}}\,
\Gamma_{f_0} (s_{\pi\pi}) }+ R_0(s_{\pi\pi})~,
\nonumber \\
F^{\rm scal}_2(s_{\pi\pi}) &=& R_2(s_{\pi\pi})~.
\label{FI}
\eea 

For $s_{\pi\pi}$ close to the  $f_0(980)$ pole, 
the leading elastic and inelastic re-scattering effects 
are automatically included in the scalar resonance
propagator  \cite{flatte}. 
In particular, considering only two-body 
intermediate states ($\pi\pi$ and $K \bar K$) and 
defining 
\be
\Sigma_{f_0}(s) = i\sqrt{s}\Gamma_{f_0}(s) \qquad {\rm and} 
\qquad  v_i (s) = (s/4-M_{i}^2)^{1/2}~,
\ee
the effective energy-dependent width
assumes the form 
\bea
 4M_\pi^2 \le s < 4M_{K^\pm}^2\  && \Sigma_{f_0}(s)=\frac{1}{8\pi\sqrt{s}}
\left\{ -(g^{f_0}_{KK})^2 \Big[ |v_{K^+}(s)|+|v_{K^0}(s)| \Big]+ 
i (g^{f_0}_{\pi\pi})^2 v_\pi(s) \right\}~, \nonumber\\
4M_{K^\pm}^2 \le s < 4M_{K^0}^2 && \Sigma_{f_0}(s)=\frac{1}{8\pi\sqrt{s}}\left\{ (g^{f_0}_{KK})^2
\Big[ -|v_{K^0}(s)|+ i v_{K^+}(s) \Big] +i (g^{f_0}_{\pi\pi})^2 v_\pi(s) \right\}~, \no \\
s \ge 4M_{K^0}^2 && \Sigma_{f_0}(s)= \frac{i}{8\pi\sqrt{s}} \left\{
 (g^{f_0}_{KK})^2 \Big[v_{K^+}(s)+v_{K^0}(s)\Big] + (g^{f_0}_{\pi\pi})^2 v_\pi(s) \right\}. 
\eea

The key observation which allows to determine the 
absorptive parts of the form factors in the low $s_{\pi\pi}$ region 
is the Fermi-Watson theorem \cite{FW}. This implies 
that below the inelastic threshold the phases 
of the two $F^{\rm scal}_{I}(s_{\pi\pi})$ 
coincide with the $\pi\pi$ $S$-wave elastic phases.
\TABLE[r]{
\begin{tabular}{|c|c|c|}
\hline 
       &  $a_I$  &   $b_I$  
\\  \hline
$I=0$  &  $0.220 \pm 0.005$  &   $0.275 \pm 0.009$  
\\  \hline
$I=2$  &  $0.045 \pm 0.001$  &   $0.081 \pm 0.002$  
\\  \hline
\end{tabular}
\caption{\label{tab:delta} Numerical values for the low-energy parameters 
of $S$-wave $\pi\pi$  phases from Ref.~\cite{colangelo}. } }
\noindent
The low-energy structure of ${\pi\pi}$ phase 
shifts is known very precisely \cite{colangelo}. 
In the $S$--wave channels we can write
$$
\tan[\delta_I(s)] \approx \frac{2 v_\pi(s)}{\sqrt{s}}\left(a_I+b_I\frac{s-s_0}{s_0}\right)~,
$$
with $s_0=4M_\pi^2$, and the $a_I$ and $b_I$ reported in Table~\ref{tab:delta}.

In the $I=2$ case, where the resonance term is absent, 
the Fermi-Watson constraint applies directly to the 
non-resonant term $R_2(s_{\pi\pi})$. Generalizing the polynomial 
expansion in Eq.~(\ref{eq:par1}), this implies 
\bea
R_2(s_{\pi\pi})=  \left(
\frac{\gamma_0 }{M_\Phi^2} +\frac{\gamma_1}{M_\Phi^4} ( s_{\pi\pi} -  M_{f_0}^2)
+ \cO[ ( s_{\pi\pi} -  M_{f_0}^2)^{2} ]\right)
e^{i\delta_2(s_{\pi\pi})}~,
\label{i=2-ff-phase}
\eea
where the $\gamma_{i}$ are real parameters, as implicitly assumed 
for all the effective couplings 
so far introduced ($\gvsg$, $\gvpg$, $\gs$ \ldots).

The condition to be imposed on the  $I=0$ term  is slightly 
more complicated since we cannot ignore the phase shift induced 
by the $f_0$ propagator. Given that below the inelastic threshold 
all phase shifts are proportional to the pion velocity 
$v_\pi (s_{\pi\pi})$, a convenient decomposition for 
$R_0(s_{\pi\pi})$ is 
\be
R_0(s_{\pi\pi}) =
\frac{\alpha_0}{M_\Phi^2} e^{i\beta_0 \frac{v_\pi (s_{\pi\pi})}{M_\Phi}}   
+\frac{\alpha_1}{M_\Phi^4} e^{i\beta_1 \frac{v_\pi (s_{\pi\pi})}{M_\Phi}}  (s_{\pi\pi} -  M_{f_0}^2)
+ \cO[ (s_{\pi\pi} -  M_{f_0}^2)^{2} ]~,
\label{i=0-ff}
\ee
where $\alpha_{i}$ and $\beta_{i}$ are real parameters.
Imposing the Fermi-Watson constraint allows to determine the $\beta_i$ 
in terms of the $\alpha_i$ and the parameters of $\delta_0(s_{\pi\pi})$. Indeed, expanding the 
phase of the full form factor in powers of $(s_{\pi\pi}-s_0)$, up to second order,
leads to the following two conditions:  
\be    
\beta_0 = \frac{M_\Phi^3}{\alpha_0}\left[ \xi(s_0)
- (s_0-M_{f_0}^2) \xi^\prime (s_0) \right]~, \qquad 
\beta_1 = \frac{M_\Phi^5}{\alpha_1} \xi^\prime (s_0)~,
\ee
where 
\bea
\xi(s) &\!=\!&
\frac{(\gpp)^3g^\phi_{f\gamma}/(8\pi\sqrt{s})}
{(s-M_{f_0}^2+\re[\Sigma_{f_0}(s)])^2+\im[\Sigma_{f_0}(s)]^2}+\frac{2}{\sqrt{s}}\left(
a_0^0+b_0^0\frac{s-s_0}{s_0}
\right)\times \no  \\
&& \times\left[
\frac{\gpp g^\phi_{f\gamma}(s-M_{f_0}^2+\re[\Sigma_{f_0}(s)])}
{(s-M_{f_0}^2+\re[\Sigma_{f_0}(s)])^2+\im[\Sigma_{f_0}(s)]^2}+
\frac{\alpha_0}{M_\Phi^2}+\frac{\alpha_1}{M_\Phi^4}(s-M_{f_0}^2)\right]
\eea
and $\xi^\prime (s)=d\xi/ds$.  

Proceeding in a similar way, this method can be generalized 
to include non-$\Phi$ contributions and/or the explicit 
pole structure of the $f_0(600)$ and/or 
additional polynomial terms in the Taylor expansion.

\section{Summary}
The procedure we propose for a general un-biased
analysis of the $e^+ e^- \to P_1 P_2 \gamma$ 
cross sections can be summarized as follows:
\begin{itemize}
\item{} According to Eqs.~(\ref{evme})--(\ref{eq:general}),
the Born (single-photon) cross sections 
are expressed in terms of the three Lorentz-invariant form factors $A_{1-3}$.
\item{} The three form factors are decomposed as in 
(\ref{eq:A_full}) in terms of a FSR component 
($A_{1,2}^{\rm FSR}$) and a two leading 
SD components ($A_{1}^{\rm scal}$ and $A_{1-3}^{\rm vect}$).
The FSR component, which is present only in 
the $|\pi^+\pi^- \gamma \rangle$ case,
is fully determined by the electromagnetic pion form factor. 
\item{} The vector SD component can be 
parameterized as in Eq.~(\ref{eq:A_vect}) 
in terms of the effective couplings  $F_V$,  
$\gvpg$, and $\gvpv$. Here one 
could use the reference values in 
Table~\ref{tab:vmd} as starting point
of the fit, and eventually improve 
the determination of some of these couplings 
(especially by means of $| \pi^0\pi^0 \gamma\rangle$ data,
where the vector component is dominant).
\item{} For the scalar SD component we propose 
a parameterization of type (\ref{eq:A1_gen})
with a resonant part (with or without the 
$f_0(600)$ pole) and a polynomial non-resonant term
(with one or two free parameters for each channel).
In the two $|\pi\pi \gamma\rangle$ channels the  
parameterization of the non-resonant part can be 
improved with the inclusion of appropriate 
re-scattering phases, as shown in Eq.~(\ref{i=0-ff}). 
These do not involve additional free parameters since 
the $\beta_i$ in Eq.~(\ref{i=0-ff}) can be 
determined using the precise low-energy 
constraints on $\pi\pi$ phase shifts,
as outlined in Section~\ref{sect:rescattering}.
An important consistency check of the whole approach
is obtained by the combined fit of the  
two $|\pi\pi \gamma\rangle$ channels,
which should satisfy the isospin decomposition 
(\ref{iso-dec}).
\end{itemize}

\noindent 
Recently, this procedure has been employed in the 
analysis of KLOE data on $e^+ e^- \to \pi^+\pi^- \gamma$~\cite{KLOEpp}
with satisfactory results. A more significant 
test of the method should be possible in the 
near future, with the combined analysis of 
high-statistics data on both 
 $e^+ e^- \to \pi^+\pi^- \gamma$ and
$e^+ e^- \to \pi^0\pi^0 \gamma$.

\subsection*{Acknowledgments}
We thank Cesare Bini, Simona Giovannella, and Stefano 
Miscetti for useful discussions about KLOE data.
This work is partially supported by IHP-RTN, 
EC contract No.\ HPRN-CT-2002-00311 (EURIDICE).

\end{document}